# On the statistical mechanics of life: Schrödinger revisited

Kate Jeffery[a], Robert Pollack[b], Carlo Rovelli[c]

[a]*Institute of Behavioural Neuroscience, University College, London WC1H 0AP, UK. k.jeffery@ucl.ac.uk*
[b]*Biological Sciences, Columbia University 749 Mudd, Mailcode 2419, New York NY 10027, USA. pollack@columbia.edu*
[c]*Centre de Physique Théorique, Aix-Marseille University, Université de Toulon, CNRS, Marseille, France; and Perimeter Institute, 31 Caroline Street North, Waterloo, N2L 2Y5, Ontario, Canada. rovelli@cpt.univ-mrs.fr.*
(Dated: August 12, 2019)

**Abstract**

We study the statistical underpinnings of life. We question some common assumptions about the thermodynamics of life and illustrate how, contrary to widespread belief, even in a closed system entropy growth can accompany an *increase* in macroscopic order. We consider viewing metabolism in living things as microscopic variables directly driven by the second law of thermodynamics, while viewing the macroscopic variables of structure, complexity and homeostasis as mechanisms that are entropically favored because they open channels for entropy to grow via metabolism. This perspective reverses the conventional relation between structure and metabolism, by emphasizing the role of structure for metabolism rather than the other way around. Structure extends in time, preserving information along generations, particularly in the genetic code, but also in human culture. We also consider why the increase in order/complexity over time is often stepwise and sometimes collapses catastrophically. We point out the relevance of the notions of metastable states and channels between these, which are discovered by random motion of the system and lead it into ever-larger regions of the phase space, driven by thermodynamics. We note that such changes in state can lead to either increase or decrease in order; and sometimes to complete collapse, as in biological extinction. Finally, we comment on the implications of these dynamics for the future of humanity.

## I. INTRODUCTION

From the perspective of physics, the early appearance of life on Earth and its tenacious resilience over billions of years implies that life must be an entropically favored phenomenon.

Life and its evolution are time-oriented, irreversible phenomena that have produced a steady increase in complexity over billions of years. The second law of thermodynamics, according to which entropy increases in irreversible processes, is the only fundamental law in physics that distinguishes the past from the future. Therefore this law, and its statistical underpinning, offer the only physical principle that can govern a macroscopic irreversible phenomenon – and yet, increasing entropy is commonly associated to decreasing order, making life's increase in complexity a challenge to explain. Life take place in regimes which are far from those studied by classical equilibrium thermodynamics, but also from those captured by the past and current attempts to describe non-equilibrium thermodynamics (see for instance Prigogine 1967, Schnakenberg 1976, Kauffman 1993, Hill 2004, Kleidon and Lorenz 2005, Bertini, De Sole, Gabrielli, Jona-Lasinio and Landim 2017, Polettini and Esposito 2019, and references therein). Here we suggest a way in which the statistical logic underpinning the second law of thermodynamics can directly drive life and evolution towards order and complexity.

We begin by recalling a number of potential confusions surrounding the relations between entropy, order, complexity, probability, and life, which often mislead the discussion on the statistical physics of life. We recall in particular that formation of order and structure driven by entropy increase are ubiquitous in the universe: cosmological formation of galaxies and stars from the initial uniform matter distribution are examples. In these, and many other cases, elaborate structures form—not against statistics, but driven by the statistical logic of the second law. We then point out some specific notions that can help in disentangling the complex relation between life and physics, in the regime far from thermodynamic equilibrium where life operates. Among these are the notions of macroscopic order providing a conduit for entropy to increase, metastable states, random motion within these states and channels among such states.

This leads us to a perspective on the possible statistical underpinning of life, whereby life is not an improbable "fight against entropy" —as Erwin Schrödinger famously put it in his adventure into biology (Schrödinger 1944)— but is rather a statistically favored process directly driven by entropy growth, in which movement of a system within a space of available states leads it to discover —and traverse — channels between metastable states. We also discuss the different roles that the notion of information plays in this context. The perspective we develop builds upon the numerous previous efforts towards understanding the statistical underpinning of life (see Ashby 1947, Nicolis and Prigogine 1977, Haken 1983, Kauffman 1993, Depew and Weber 1995, Hill 2004, Bialek 2012, England 2013, Ramstead, Badcock and Friston 2018, and especially Perunov, Marsland and England 2016).

We also briefly comment on the specificity of our own species in this regard, and the existential risks it faces, which are manifest when considered from this perspective.

## II. ENTROPY AND ORDER

Let us begin by dispelling some common confusion. Central to this discussion is the notion of order: that is, of correlations in time and space that mean that particles or phenomena in the universe are not randomly dispersed, but rather have spatial or temporal structure. It is often assumed that any increase in order must imply a decrease in entropy. The strict equivalence *order=low entropy* is a mistake. It is a persistent prejudice that misleads the efforts to understand the physics of life.

The key point, which we illustrate in detail below, is that *macroscopic* order can bring energy from macroscopic to microscopic variables, increasing entropy as it does so. This can happen not just when non-equilibrium driving bring about novel patterns of complex organization in thermally fluctuating many-body systems or 'active matter' mixtures (Cross and Hohenberg 1993, Gollub and Langer 1999, Schaller, Weber, Semmrich, Frey and Bausch 2010, Sanchez, Chen, DeCamp, Heymann, and Dogic, 2012), but also in simple isolated systems.

There are familiar cases where an increase in entropy does amount to an increase in disorder. If we put some wine in a glass of water, for example, the wine is initially concentrated: this is an ordered configuration because the wine is collected in the same region, rather than dispersed: the locations of the molecules are somewhat correlated. If we let the system evolve freely, the wine mixes with the water: the order is lost. This is an irreversible process and entropy increases. But consider the case where, instead of wine, we put some oil into the water and mix. Let the system evolve freely: the oil *separates* from the water and order is generated. But this is *also* an irreversible process, and again entropy increases. Therefore increase of entropy generates disorder in some cases, but generates order in others.

There are very many similar examples in which entropy increase generates order and structure. Since this is crucial for understanding structure formation in life, let us consider a few further examples.

Consider an isolated box where heavy (subject to gravity) balls bounce against the floor, walls and ceiling, and against one another. In the idealized situation where bounces are perfectly elastic, there is no dissipation, and energy conservation implies that the balls continue to bounce forever: the dynamics is reversible. In a realistic situation there is dissipation at each bounce: hence heat is generated, and hence entropy plays a role. The macroscopic dynamics becomes irreversible, and the balls end up motionless, *in an orderly arrangement on the floor*. Their mechanical energy has been dissipated into heat, increasing entropy. Hence, entropy increase accompanies progression from a macroscopically disordered situation where the balls were all over the box, to a macroscopically ordered situation where they are lying in an orderly way on the floor.

This generalizes. For instance, it is entropy increase that separates air, water and rock on Earth, and is responsible for the existence of the sea, which stays above the rocks and below the atmosphere. Without dissipation, and hence entropy increase, water and air molecules would continue to mix. The common orderly arrangement on the Earth's surface, whereby the atmosphere is above the sea and the sea is above the rock, depends on the second law of thermodynamics.

A last example. Consider air saturated with water vapor, and lower its temperature below the water-freezing point. Water vapor condenses into snowflakes, or frost forms on walls (depending on specific conditions, such as the presence of nucleating factors). Observed with a magnifying glass, snowflakes and frost display magnificent elaborated architectural structures, which we identify with order. The formation of snow and frost is an irreversible process driven by entropy increase: once again, entropy increase generates order: an elegant and elaborated form of order, in this case.

The general conclusion is clear: entropy increase generates macroscopic disorder in some cases, but macroscopic order of various kinds in others.

The relevance of this observation for biology is that the widespread idea that life is "a local fight against entropy", namely a trick to keep entropy locally low[1], is misleading. The idea follows from the (correct) evidence of important amount of order in life (life is a self-organizing process of propagating increasing complexity) and from the (wrong) prejudice that order and complexity necessarily imply low entropy.

---

[1] "The essential thing in metabolism is that the organism succeeds in freeing itself from all the entropy it cannot help producing while alive." Schrödinger, E., "What is Life," 1944, chapter 6, Order, Disorder, and Entropy. To be sure,

Since higher entropy is related to higher probability, this idea might lead to the misleading conclusion that life must be naturally improbable. From the perspective of physics, as well as to a casual observer, such a conclusion is absurd: the improbability associated with lowering of entropy is dramatically strong: if there really were anything entropically improbable in life, life would not have happened on Earth. And if it had, it would have been soon demolished by thermal fluctuations. The (early) appearance and the tenacious resilience of life on Earth over 4 billion years makes it obvious that life cannot be a "struggle against entropy" in any sense: it can only be an entropically favored phenomenon. The inexorable progression of life from simpler and less ordered to complex and more ordered forms suggests that entropy and complexification must be related. It is the aim of this paper to discuss how this can be so.

### III. ENTROPY AND STATISTICS

Since the second law of thermodynamics is the *only* fundamental law that distinguishes the past from the future, the physical basis of any irreversible phenomenon is this law, and nothing else. Before discussing in which sense the second law drives life, we briefly review here the best current understanding of the statistical underpinning of this subtle law, largely due to Boltzmann.

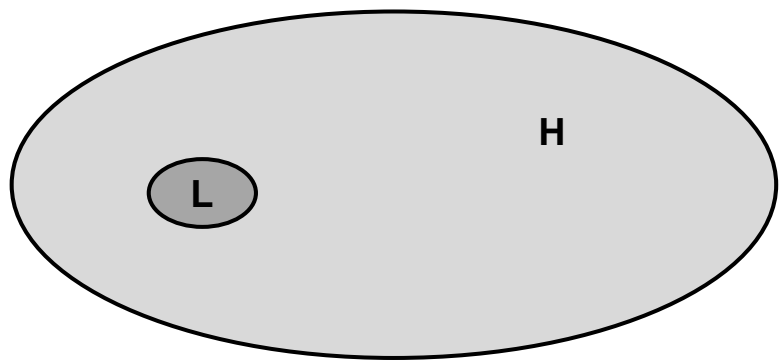

Figure 1: The intuitive understanding of the logic of the second law. The space in the picture represents all possible states of a system. If (i) there is a variable that has value *L* in a small region and value *H* in a large region, and if (ii) the evolution *begins* in *L*, then a it is likely to end up in *H*. The converse is not true: a generic evolution that *begins* in *H* remains in *H*. Hence the system evolves irreversibly from *L* to *H* but not vice versa.

The second law of thermodynamics concerns the evolution of *certain variables* of a system. This is subtle, but important for what follows. It applies to only some variables, possibly pertaining to the (relatively) large-scale properties of the system, called "macroscopic variables", which can be numerous, but are still few in number compared to the number of degrees of freedom in the system. Entropy is a function of these variables that measures the number of states (or the volume of *phase space* – the space of possible states) where the variables have a given value. A "macroscopic state" is a set of values of

---

Schrödinger's book is subtle, and the misconception we refer to (that life acts by holding on locally against thermodynamic dissipation) is not the main idea of the book. Schrödinger obviously did not think that life is at odds with what he called the "probability mechanism", the second law. But he thought that something else holds in life but not in non-life. Another mechanism, responsible for turning the microscopic order encased in a single DNA molecule into the macroscopic order which we identify as a living organism—"order from order", as he puts it. Contrary to this, we suggest here that no new principle is need: life is driven directly by the second law, via a complex cascade through channels between metastable states opened by structure.

these macroscopic variables, and it corresponds to an ensemble of "microscopic" states, whose size is measured by entropy.

Notice that with this statistical definition (due to Boltzmann) the notion of entropy is defined for any value of the macroscopic variables, and not necessarily at equilibrium. The second law applies when in the past entropy is lower than its maximum. It states in particular that entropy increases in irreversible phenomena.

The distinction between work and heat (two forms of energy), which lies at the foundation of classical thermodynamics, is based on the macro/micro distinction as well: work is energy in macroscopic variables, while heat is kinetic energy in the microscopic variables. Importantly, macroscopic variables here are not just pressure, temperature and volume, but *all* the variables that enter in the description of a system such as those describing the position of distinguishable objects, the structure of an organism, the instantaneous chemical composition inside its cells, and so on. 'Macroscopic' variables can include, for instance, 'small' variables such as DNA nucleotides sequences, as we shall see below. Microscopic variables, on the other hand, are positions and velocities of all the individual molecules in the system. Energy available to do work is often called *free energy*, and is a measure of how much entropy the system can still gain (i.e., how far away it is from thermal equilibrium).

Let's recall why entropy increases, if it were low to start with. A low-entropy state occupies a tiny region in the space of states (see L in Fig. 1). A system starting from one of the rarer low entropy states in the state space will therefore, under the laws of probability, spontaneously tend to move out into a more common higher entropy state: the reverse is statistically much less likely. This picture illustrates also why processes evolve towards maximal entropy, if they can: they tend to flow to the largest reachable region, simply because the majority of the points where they can randomly arrive lie in the large region. This drive towards maximal entropy is the 'reason' for all irreversible processes. This is what is meant by the metaphor of the system that "wants to increase its entropy". This is the logic underpinning the second law.[2] It is not only relevant for the behavior of the systems at or near equilibrium states, which is quantitatively accounted for by equilibrium thermodynamics, but for any irreversible process in nature as well.

We can now revisit why entropy increase can generate order. Consider the simplest case: the bouncing balls collecting on the bottom of the box. Why does this increase entropy? The reason is to do with Boltzmann's characterization of entropy as being the number of microstates that correspond to a given macrostate. The number of *macroscopic* configurations where the balls lie on the bottom is smaller than the number of macroscopic configurations where the balls bounce around, and hence there is greater order. But the number of *microscopic* configurations where the balls lie on the bottom is much higher than the number of microscopic configurations where the balls bounce around – because when dissipation transforms mechanical energy into heat (that is: moves energy from macroscopic to microscopic variables), the individual molecules of the balls access a larger region of their phase space, previously inaccessible by energy conservation, because they can have a wider range of velocities. Hence there are far more microscopic ways of having slightly warmer balls lying in order on the floor than colder balls bouncing, and hence there is also higher entropy. Entropy increases by ordering the

---

[2] The second law has also a mysterious side. The mystery is why entropy was low to start with, in our universe. This is mysterious because we understand entropy increase in terms of genericity (it increases *generically*, namely for most microstates); but the initial state of (certain macroscopic variables describing) the universe was low-entropy, which is a non-genericity assumption (most microstates did not have this property). Here we do not discuss this mysterious aspect of entropy: we take as a fact of the universe in which we live that there are macroscopic variables whose entropy was low in the past. For a tentative interpretation of this fact, not related to what we do here, see (Rovelli 2016a).

*macroscopic* position of the balls, and transferring energy to the thermal agitation of the molecules (*'disordering' them*).

The water/oil dynamics is the same mechanism. Any mechanism for which an ordered macroscopic configuration brings energy from macroscopic to microscopic variables increases the entropy, while increasing macroscopic order. In the case of the balls, it is gravity that does the job; in the case of oil it is intermolecular forces in oil. In general, many kinds of chemical processes can do the same.

This is relevant for life. Biological structure is a form of order, related to metabolism, which itself is an entropy-increasing process. Structure formation driven by increase of entropy is a common physical phenomenon. Cosmological formation of galaxies and stars from the initial uniform matter distribution are other well understood examples. In these, and many other cases, elaborate structures form—not against statistics, but driven by the statistical logic of the second law.

### IV. CORRELATION AND INFORMATION

Let us be a bit more quantitative. Consider a physical system with a large number of degrees of freedom, with phase space $\Gamma$. To simplify the mathematics, assume that the set of its possible states is discrete. This does not affect the general logic. Denote individual states as $s \in \Gamma$. Consider a set of macroscopic variables $a_n$ (with discrete values as well for simplicity), functions on $\Gamma$. For any set of values $a_n$ define the "Boltzmann entropy" (so denoted although this formula was first written by Planck: Jona-Lasinio 2015)

$$S(a_n) = k \log W(a_n) \tag{1}$$

where $k$ is the Boltzmann constant and $W(a_n)$ the number of states where the variables have values $a_n$, that is

$$W(a_n) = \sum_s \prod_n \delta(a_n(s) - a_n) \tag{2}$$

where $\delta(x) = 1$ if $x = 0$ and $\delta(x) = 0$ otherwise. This defines entropy.

There is no canonical definition of order. We take here 'order' to be the property of a state that allows some of its aspects to be predicted from others. The oil on water is ordered because once you know where some of the oil is you can guess at the location of the rest. A snowflake is ordered because if you see a picture of half of it, you can have a good guess about the missing half. The molecules of a gas are *dis*ordered, because knowing the position of a few molecules tells you little about the position of the others. Ordered states are thus states obeying constraints between variables associated to different degrees of freedom.

To make this precise, consider two degrees of freedom of the system, described by the variables $a$ and $b$ respectively. Let $N_a$ and $N_b$ be the number of possible values that the variables $a$ and $b$ can respectively take. Let $N_{ab}$ be the number of possible values that the couple *(a, b)* can take. If there are no constraints, clearly $N_{ab} = N_a \times N_b$. But $N_{ab}$ can be smaller than $N_a \times N_b$ if a physical constraint is in place. For instance, each end of a bar of magnetic iron can be of North or South polarity, so that $N_a = N_b = 2$, but physics implies that $a$ and $b$ are opposite: hence only two combinations of the couple *(a, b)* are allowed, namely *(N, S)* and *(S, N)*, so that $N_{ab} = 2 < N_a \times N_a = 4$. This leads us a formal definition of order: we say that there is *order* if

$$N_{ab} < N_a \times N_b.$$

(3)

The existence of order allows us to know something about the value of *a* if we know the value of *b*.

The amount of order – defined in this way – is quantitatively measured by Shannon's relative information *I* (Shannon 1948). This is defined, in the case at hand, by

$$I = \ln_2 N_a + \ln_2 N_b - \ln_2 N_{ab}. \tag{4}$$

This expression can be generalized for many variables, continuous variables, probabilistic variables and so on; here we do not do so, since we need only the basic idea.

Notice the appearance of the notion of *information*. Eq. 4 (and its generalizations) is a very general definition of information, which amounts to a measure of correlation. The idea is that we can say that one end of the bar "has information" about the other, in the sense that by knowing the polarity of one end we know the polarity of the other. The word "information" is used with a vast variety of different meanings in different contexts. This definition of information is extremely general and based only on physical correlation. Later we will refer also to a more restricted definition ("relevant information").

These definitions already allow us to render with quantitative precision the observation that entropy and order can go together. When the balls dissipate their kinetic energy, the oil separates from the water, or snowflakes form, Boltzmann entropy (1) clearly grows, but Shannon's relative information (4), which measures order, grows as well, because the macroscopic variables measuring the positions of the balls, the oil drops or the ice particle become correlated. In this case, growing entropy produces order.

Quantitatively, if the total kinetic energy of the *n* balls was initially *E*, and the temperature *T*, the order of magnitude of the increase in entropy due to the dissipation of this energy is

$$\Delta S \sim \frac{E}{T} \tag{5}$$

which means that after dissipation the volume of the phase space accessible by the micro-states increases by the factor

$$\rho^{micro} \equiv \frac{V^{micro}_{final}}{V^{micro}_{initial}} \sim e^{\frac{\Delta S}{k}} \sim e^{\frac{E}{kT}}. \tag{6}$$

The volume of the phase space accessible by the macrostates which is lost by the balls in ordering themselves on their gravitational ground states, on the other hand, can be estimated comparing the free particle phase space volume $V \sim E^{\frac{3}{2}n}$ with that accessible at thermal equilibrium

$$\rho^{macro} \equiv \frac{V^{macro}_{final}}{V^{macro}_{initial}} \sim \left(\frac{E}{kT}\right)^{\frac{3}{2}n} \tag{7}$$

If the macroscopic motion of the masses were not just due to thermal fluctuations $E \sim kT$ and for a finite number of particles

$$\rho^{micro} \gg \rho^{macro}, \tag{8}$$

because $e^x \gg x^n$ for any finite *n* and very large *x*.

That is: the macroscopic variables get ordered by disordering the microscopic variables. Entropy can lead to an increase in macroscopic order.

So far, we have only seen that entropy *can* drive up order. But why *does* it do so, and to the extraordinary degree that characterizes biological matter? Why does evolution progress in the direction of increasing complexity over time? We need several steps to get there. The first is the notion of metastable state.

## V.  METASTABLE STATES

An ideal gas expanding in a box reaches equilibrium —namely its maximum entropy state— quite rapidly. The equilibrium state is stable: that is, the macroscopic variables do not change any further. In nature, however, this is rare: typical systems equilibrate to metastable states, not to stable states. Metastable states appear to be stable possibly for long periods, but they are not at equilibrium: that is, they do not have maximal entropy.

Metastable states are ubiquitous. For instance, a pile of wood in a room full of air seems very stable. But it is not, as any owner of a storage facility can tell you. It can burn. Combustion is an irreversible process that takes the metastable state of the wood to a much higher entropy state. The room was in a metastable state: a state that can remain stable for very long time, but its entropy can still increase substantially. How can a system be so thermally stable if it is possible for its entropy to grow?

The answer is that 'possible' is different from 'likely'. Metastable states can be understood in terms of their statistical mechanics as regions of phase space that have less volume than truly stable states, but that are connected to these by narrow gaps that a random motion in phase space has difficulty finding. See Figure 2 for an intuitive explanation.

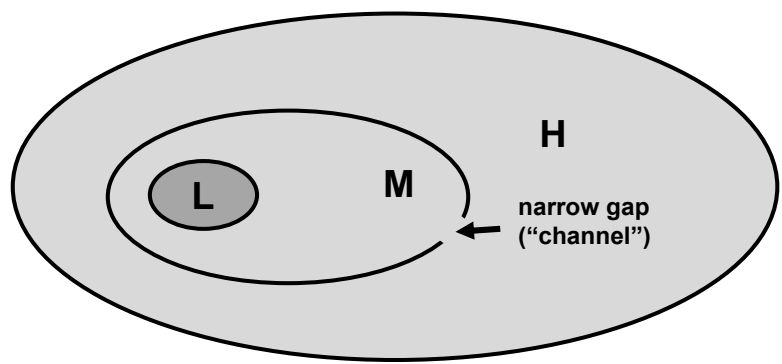

Figure 2: Intuitive representation of metastable states: the metastable state is the region *M* of phase space in a system for which the dynamics cannot cross the boundary of *M* except through a very narrow gap. A microstate leaving *L* and moving randomly will remain long trapped in the region *M* (the metastable state) before accidentally finding its way out towards the stable state *H*.

Most if not all the states that we call "stable" are actually metastable, with lifetimes longer than the observation time.

The reason for this ubiquity is the complexity of the dynamics. The physical dynamics of Nature is governed by (non-linear) equations that have a very large number of variables and a very rich phase-space structure. This can include regions where the system can easily move ergodically, connected to one another by narrow gaps, which require long times for the micro-state to randomly find. These narrow gaps are called "channels".

A related notion commonly used in biology is that of attractor states, in which a system tends to remain within (is attracted to) a small region of the phase space but can switch abruptly to a different, also-stable state. An example is the rapid reorganization of the neural code for place when environmental variables are changed: the system ignores small changes but suddenly massively reorganizes when the change is sufficiently large (Jeffery, 2011).

## VI. CHANNELS

The channels that connect metastable states with higher-entropy states can be themselves opened or closed by the dynamics: either by a rare random fluctuation, or by an external action (as with the environmental change described above). A pile of dry wood in the room burns if we ignite the fire. The combustion is an irreversible process that takes the metastable state of the wood to a much higher entropy state.

A large astrophysical hydrogen cloud can remain thermodynamically relatively stable for millions of years. But there is a small gravitational instability, the Jeans instability (Jeans 1902), that can make it slowly shrink. The shrinking increases temperature. Temperature can rise to the point where the hydrogen starts fusing into helium. This is an irreversible process, which brings the hadronic matter from the metastable hydrogen state to the much higher-entropy helium state.

An interesting aspect of this example is that the opening of the channel for entropy to grow is caused by the shrinking of the cloud. An increase in macroscopic order opens a channel for a large increase of entropy. The widely different scales involved determine the long durations of these processes, which can last billions of years. This is a mechanism that plays a key role in the statistical mechanics of life.

The biochemistry of life is a network of chemical processes, each of which is individually driven by entropy increase: no chemical reaction happens unless entropy increases. But individual biochemical reactions are chemically regulated by enzymes, which (from the perspective of statistical mechanics) are dynamical mechanisms that open and close channels for entropy to increase. In the absence of the relevant enzyme, two biochemical reagents remain inert. They are trapped in a metastable state of lower entropy. The presence of the enzyme opens a dynamical path to a higher entropy state: the reaction happens, driven by the second law of thermodynamics.

There is here a key hint: an enzyme is a very *ordered* arrangement of atoms. An enzyme is thus a manifestation of order. The presence of this order is the factor that opens the channel for the reaction to happen and entropy to increase. That is: order can open channels for entropy to grow.

## VII. THE PHASE SPACE OF BIOLOGICAL SYSTEMS AND PERCOLATION IN IT

A biological system is typically an ensemble of atoms that include oxygen, hydrogen, carbon, and a few other elements. Naively, given so few elements one might think that the structure of the corresponding phase space should be relatively simple, but obviously this is not the case, in particular because of the extraordinary complexity of carbon chemistry, which arises from its ability to polymerize in all three dimensions. This opens the space for the extreme richness generated by combinatorics. The complexity of carbon chemistry is not a product of life, it is the aspect of the structure of the relevant physical state space that underpins it.

The dynamics of the interactions between organic chemicals is even more complex, as the interactions influence one another in all sort of manners. Therefore the structure of the relevant microscopic phase space is extremely complex: it is full of possible metastable states and channels between them. Life is a percolation among these channels (crf.: Ramstead, Badcock and Friston 2018). Not only is it far from equilibrium, it is also very far from having explored the full space of this complexity: for instance, only a minimal fraction of all possible proteins has been explored by life so far.

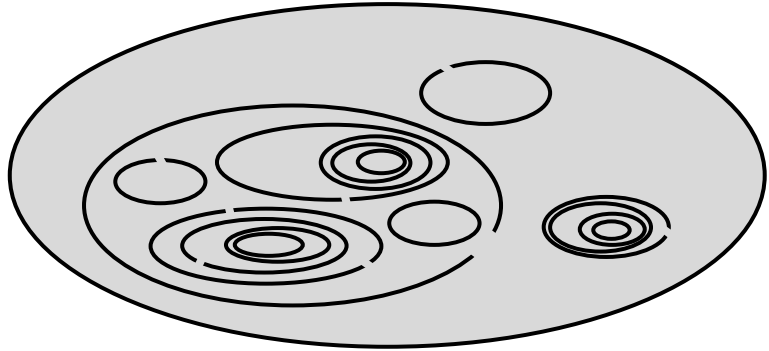

Figure 3: Intuitive (oversimplified) representation of the complex phase space of living physical system: extremely numerous metastable state regions are nested within one another.

What drives the percolation? The statistics of the second law of thermodynamics, of course, because nothing else could. That is: the simple tendency of micro-states to migrate randomly towards larger regions of phase space.

For this to happen we need low entropy to drive the process. Of course the Sun is a very rich source of low entropy (free energy), because the photons coming from the sun have far less entropy than those leaving Earth, which are cooler. Therefore the biosphere is constantly bathing in a large wealth of free energy, which drives the biosphere processes. Each single process is driven by entropy increase, whenever it is allowed by channels opened by the dynamical processes permitted by ordered structures. The point about understanding life's statistical properties is not how biochemical reactions happen, of course: it is to understand how the regulating action of the enzymes (and the rest of the orderly structure of life) is itself entropically favored. To get there, we still need to sharpen our statistical picture.

## VIII. CORRELATIONS IN TIME AND INFORMATION

Above, we have defined a notion of order in space. At a given moment, order is the existence of correlations between different variables of the system. There is certainly order of this type in biology, in particular in the orderly arrangement of biological structures. The most relevant biological order, however, is of a different kind.

Life is a process, not a state. We can only hope to understand it as a process in time: not as a state at a fixed time. As a state at a fixed time, it looks surprising. It is by looking at it as a long term process that we can understand it. This is a common pattern in science: the structure of the solar system or the structure of atoms were understood when instead of trying to figure out their instantaneous structure, scientists shifted their attention to their dynamics: the structure of the atom is understood by studying how its electrons *move*.

Let us therefore think diachronically rather that synchronically; that is, let's consider the long term temporal evolution of the systems, rather than their instantaneous state.

We need two concepts. Instead of the phase space of the system, we should focus on the space of the *motions* of the system: namely the space of the possible histories (past and future) of the system. There is a strict relation between the two, because if we fix a time, each microstate at this time uniquely determines a motion and vice versa. (Here we disregard quantum uncertainty.) We can extend the notion of coarse graining to the space of motions: macroscopic motions are families of ("microscopic") motions that are not distinguished by the macroscopic variables.

Since we are using the second law of thermodynamics and assuming initial low entropy, we restrict our discussion to those motions that start in a suitable low-entropy region. We are interested in the properties of a generic macroscopic motion among these, with regard to order and structure formation.

The second notion we need is correlations in time rather than in space. These can be defined as follows. Given a variable $a$ that takes the value $a(t)$, we say that there is order if there is a correlation (in the sense defined above) between $a(t_1)$ and $a(t_2)$, where $t_1$ and $t_2$ are different times.

Life is then first and foremost characterized by a spectacular amount of correlation across time.

Recalling that correlation is information, this can be expressed in informational terms: life is characterized by a spectacular amount of preservation of information across billions of years. One key holder of this information is of course the DNA molecule and the information it encodes.

Using the precise definition of information given above (Shannon's relative information as physical correlation), we can distinguish three distinct senses in which DNA molecules carry information.

(i) Each single strand of a double-stranded DNA is the template of the other: given one, we can predict the other: hence it is correlated with it; hence it has information about it. (A single strand of DNA in isolation has no information, as defined above: it is just a sequence, as unique as any other sequence.) The double strand has relative information because each strand has information about the other. This is key for reproduction.

(ii) DNA encodes proteins and is therefore correlated with the proteins produced: in this sense, it has information about the proteins structuring the organism.

(iii) What mostly concerns us here is the *third* sense in which DNA molecules carry information: the entire molecule has information because it is reproduced across time - it has correlations with the past and the future. A double-stranded DNA and its descendants carry information for billions of years, through semi-conservative replication (that is, replication for which each copy comprises half its parent molecule and half a newly assembled duplicate). The macroscopic histories that contribute to biology are histories characterized by *huge amount of temporal correlation*: information across very long time periods.

We are not referring here to the lifetime of a single molecule, of course. We are referring to the fact that there is a correlation between the order of the sequence of the bases at some time and the order of the sequence of the bases after a long time span. This correlation is generated by the transmission of the information of a DNA sequence of base-pairs through time by semiconservative replication. The correlation is stronger within the lifetime of a species, but is also present across far longer time scales, because elementary biological mechanisms are shared by vast numbers of species. Notice also that these long-term correlations exist in spite of the mortality-linked metastability of any single DNA molecule carrying that information, and are in fact to some extent a consequence of it.

What could make this astonishing amount of order statistically favored? Only one ingredient: it favors entropy growth. And the means by which it does so: by sustaining life's metabolism.

## IX. THE STATISTICAL UNDERPINNING OF LIFE

The long-term temporal order opens channels for entropy to grow, and does so abundantly by doing so repeatedly in time. Since entropy "wants" to grow, the second law favors macroscopic histories in which this order is realized. For this to happen, the opening of the channel must be self-sustaining.

This leads us to recognize life as a phenomenon characterized by the following features:

(i) metabolism is a process that makes entropy grow, and as such it is directly entropically driven;

(ii) metabolism would not happen if it were not for the biochemical structure of living matter: this structure provides channels in the complex phase space of carbon chemistry from metastable states to higher-entropy ones;

(iii) inheritance is the process that allows the mechanics to be efficient in the long term, because total entropy continues to grow. This is possible thanks to long-interval correlations in time. These are given by the preservation of information, in particular in the DNA;

(iv) the structure that supports metabolism grows more complex over evolution because each complexity step opens new channels to higher entropy states.

The densely packaged relative information in the DNA, discussed above, determines a structure which is produced by entropy growth, but is also capable of re-opening channels for further entropy growth.

The main mechanism is of course the fact that the two strands of the DNA, which have information about each other, can separate, and an entropy-producing process can refurnish each strand with a new companion strand. The statistical physics of self-replication using the tools of non-equilibrium statistical mechanics is discussed in (England 2013.) The point we are emphasizing in this paper is that this is an entropy-producing process *that creates new structures capable of permitting a new entropy producing process by opening new channels for entropy to grow.* As Francis Crick was once heard to say: "All life, sir, is DNA's way of making more DNA". Since the process can repeat, information is carried in time, and *has* been carried in time for some four billion years.

DNA has of course information also because it correlates with the proteins it encodes. This rich relative information is structure that is not only nourished by metabolism, but more crucially *allows* metabolism, opening channels for it to happen, hence allowing entropy to grow. And it is this entropy growth that drives the system. If this is anywhere possible, there is no reason for it not to happen.[3]

In the perspective we are developing here, therefore, it is not metabolism that is responsible for the building and the homeostatic conservation of structure. It is, primarily, the other way around: metabolism happens just because it is driven by the second law of thermodynamics. Structure that is there has survived because it allows sustained metabolism: this has made it entropically favored.

Biochemical reactions increase entropy. Biomolecular systems like molecular motors or pumps, transcription and translation machinery, and other enzymatic reactions have an entropy cost (Barato and

---

[3] To be sure, we are not invoking a principle of *maximal* entropy production. We are simply observing that life is a normal process where entropy grows, which happens simply because microstates naturally move to larger regions of phase space, although in a complex manner.

Seifert 2015); as emphasized by Hoffmann (2012), they are *driven* by the disorder of the molecular storm and the second law of thermodynamics. The perspective we are suggesting here is to view these thermodynamical driving forces not just as functional for the building and preservation of stucture and information, but also, the other way around, as the primary driving force of the entire biosphere: structure and information being just ways through wich these thermal driving forces act, thanks to the opening to channels out of metastability and into a new region of the phase space.

In this respect, a very simple model of life is a candle: the wax of the candle burns because burning is an entropically favored process: it is a process that increases entropy. A candle that is not lighted is in a metastable state and does not burn. Lighting the wick opens a channel that allows the (entropy growing) process which is burning to happen; the flame sustains itself via a self-regulatory (homeostatic) feedback mechanism: if the flame is too strong it melts more wax, which suffocates it; if it fades, it consumes the melted wax, and liberates a new part of the wick, thus reinforcing itself. This is an elementary prototype of biological homeostasis. The poetic metaphor of life as a candle is therefore quite a good one in this respect. Of course a burning candle does not have the self-reproducing mechanism of life.

Notice that in all that, biological processes characteristically involve different scales: the scale of a DNA molecule is much smaller than the scale of a multicellular organism or an ecosystem, but the scale of the molecular thermal fluctuations where the free energy ends up is much smaller than the scale of a cell. Life is produced by the reciprocal interactions of processes at all these scales. It is constantly fuelled (largely) by the free energy produced by the Sun, at a still larger scale. This direct dynamical interaction between different scales is what makes life work and also what makes it harder to model simply. The full statistical underpinning of life is not captured by any of these scales alone.

An organism our size has $\sim 10^{14}$ cells, and one of our cells has $\sim 10^{14}$ atoms. The length of, say, a DNA molecule is negligible compared to the overall number of atoms in the cell. The molecular biological structures that are ordered, or correlated, define a 'macroscopic' coarse-graining in the atomic phase space; the corresponding macroscopic states of a cell are still formed by a vast number of microstates, given by the random configuration of the molecular thermal storm in which these structures bathe. Organic macromolecules formed from many atoms are ordered, as is a deck of cards boxed in proper order. But the reason such structures exists in an entropic world is simply that they are immersed in a thermal bath where the molecules are constantly refreshed by catabolism and anabolism working together, as metabolism. Metabolism, by virtue of the constant breakdown and reassembly of these fragile structures over a lifetime, is where the contribution comes to net entropy.

A cell is very far from being a deck of cards in proper order: a cell is a constantly crumbling, constantly rebuilt edifice using incoming low entropy ingredients, DNA information, and small molecules to assemble itself ; and using the same DNA information to entropically degrade the parts that don't work, like cells of the body that have died while we are alive, and like our entire bodies once we die. Mortality assures the second law is observed. Metabolism is the short-term sign of mortality.

### X. CONSEQUENCES OF VARIABILITY

Statistical systems do not remain exactly on their averages: they fluctuate. In a process like life, which involves so many different directly interacting scales, the fluctuations are far from being negligible, because small fluctuations at small scales have large effects on large scales. A small number of random DNA mutations, for instance, can give rise to a new species and may thereby subvert an entire ecological system.

Fluctuations are not an accidental aspect of life: they are at the core of its functioning. As extensively described already by Darwin (1859) in the first half of *On the Origin of Species by Means of Natural Selection*, the one key ingredient of evolution is the wide variability of the organisms and the fact that this variability constantly explores new forms. One source of this variability is "errors" in DNA replication that begin with a random error of a mis-matched base-pair and that becomes a hereditable error when the mismatch is corrected by the insertion of a proper but new base-pair. These are simply caused by random events: namely the random exploration of the microscopic phase space happening to find new channels. Most of these errors lead nowhere; some lead to biological novelty. Another source of variability is provided by the immensely vast combinatorics generated by the mixing of genes in sexual reproduction.

It is precisely these randomly generated processes that have gradually built the complexity of the biosphere. From a statistical mechanics perspective, this is just an expression of the random motion of the dynamics, which occasionally finds channels towards new entropically favored regions. The space not yet explored is still immensely larger than the one explored: it suffices to recall that the number of possible (human-length) genomes is $4^{3,000,000,000}$ when there are only $10^{80}$ particles in the universe.

At the end of the day it is therefore disorder, rather than order, that has allowed life to develop. It is *directly* the disordering power of randomness that is the ultimate designer of life.

Our name for the metastability of life is "evolution". The biosphere explores some wider and wider regions of the spectacularly rich phase space of carbon biochemistry, allowing increasing paths for entropy to grow. This is happening on Earth's surface because Earth has the right temperature and pressure to allow for complex carbon chemistry, and bathes in the low entropy of the radiation of the sun.

The entire biosphere is a very resilient process, but not a stationary one. It is formed by structures that themselves tend to have finite lifetimes at the levels of both individuals and species. The randomness of the fluctuations governing the process, and its trial-and-error mode of functioning, mean that at every level there are dead-ends: life is full of dying individuals, dying species, collapsing ecosystems, mass extinctions, and the like.

In other words: the fact that life is entropically favored does not mean that there is anything necessarily stable in biology: it is rather the other way around: life is based on metastability, not on stability, and metastable states have maybe long, but ultimately finite, lifetimes. Evolution is far from being monotonic because the same variability can at times be lethal.

## XI. THE IMPROBABILITY OF LIFE REVISITED

Probability reasoning is notoriously tricky, especially when dealing with single phenomena, and the Earth's biosphere is a complex but single phenomenon. A common mistake is to confuse the probability of a class of events with the probability of an event within that class. There is nothing special in being unique, because everybody is unique. This is the lottery ambiguity. If one million tickets are sold for a lottery every week, what is the probability that one ticket wins the lottery? The question is ambiguous because it can be interpreted in two ways: either (i) as the probability that a given ticket wins the lottery, which is low (one in a million); or (ii) as the probability that there is a ticket that wins the lottery, which is very high ($\sim 1$), as every week somebody wins the lottery.

This ambiguity may misleads us in evaluating probabilities.[4] The phase space of carbon chemistry is

---

[4] This is the ambiguity behind the widespread myth of the "finely tuned initial conditions": for me to exist, an extraordinary set of improbable circumstance had to be realized. For instance, my future father had to go to that

incredibly vast and therefore there is nothing strange in the fact that we see peculiar unique structures: the vast majority of configurations are equally peculiar and unique. The relevant question is not why a particular form of complexity developed. It is why complexity developed at all, and what sustained its persistence. What allowed information to be preserved through so many millennia?

At the light of the discussion above, the improbability of life melts away. How likely is it that in the extremely complex energetic landscape of the physical world, entropy would increase by persistent channels opened by structures ordered synchronically and diachronically? Intuition says now: very likely.

At our temperature, pressure, and average chemical composition, carbon chemistry, with its symmetric three-dimensional structure allowing many different shapes to form, offers an obvious playground for nature. At different temperatures, pressures, and chemical compositions, Nature has most likely all sorts of other playgrounds to experiment with, for most of which we probably would like science and imagination to figure them out, without direct observation.

The number of ways this can happen is hard to estimate precisely because the landscape is so complex. We can only reason *a posteriori*, and take the existence of the biosphere, its resilience and its early appearance as argument for their likelihood. We are not claiming that we have computed that life is probable: but we have given arguments showing that common reasons for feeling it is improbable do not hold.

**XII. EVOLUTION AND ITS STEPS**

Szathmary and Maynard Smith have pointed out that evolution is characterized by sudden transitions in complexity (Szathmary and Maynard Smith, 1995). They outlined seven major transitions, but by taking a more fine-grained view we can identify others, collectively forming the following:

1. Replicating molecules to populations of molecules in compartments (that is, cells).
2. Unlinked replicators to chromosomes.
3. RNA as gene and enzyme to DNA and protein (the genetic code leading to protein assembly).
4. Photosynthesis, which triggered a catastrophic change in the biosphere that caused a mass extinction (the Great Oxygenation Event and associated mass extinction).
5. Prokaryotes to eukaryotes.
6. Gene regulation, which enabled cellular differentiation and the development of complex multicellular life-forms.
7. Asexual clones to sexual populations, which allowed the combinatorics discussed earlier.
8. Solitary individuals to colonies (non-reproductive castes) as we see with social insects like ants and bees.
9. Neurons, which enabled organisms to coordinate activity across their cellular agglomerations and to organise specialised behaviours such as muscle contraction: leading among other things to the Cambrian substrate revolution which again caused a mass extinction.

---

particular restaurant that particular day when my future mother was there and sit at the right angle to spot her, and so on. If I say that the occurrence of all these coincidences is so improbable that certainly the universe had to be finely initially tuned to allow my own existence, I make a logical mistake. I confuse the probability of a single winning versus somebody winning. A single history of the universe is neither probable nor improbable. In spite of numerous attempts in the literature, the universe is far too complicated for us be able to reliably compute what really would happen with different initial conditions.

10. Synaptic plasticity: increase and decrease in synaptic strength, or "re-weighti*ng*", which enabled traces of neural activity to be stored in the network as memory, and allowed organisms to use the past to predict the future.
11. Internal cognitive representations, enabling organisms to simulate reality and make predictions.
12. Development of symbolic reasoning and language in humans, enabling transmission of accumulated information across time and space.

The reason for this step structure becomes particularly transparent in the light of the statistical interpretation of life developed here, and supports it: if life is the opening of stable channels for entropy to growth, then evolution, which is a slow random exploration of its phase space, should reflect the effect of this exploration by suddenly discovering new major channels for entropy to grow. At some new discoveries, we might expect a jump in its ways of raising entropy and, hence, a novel flourishing of life

Each jump accompanied a major increase in biological diversity, and each can be understood as the abrupt acquisition of new stable pathways for entropy to grow, stabilised by the long term temporal correlations permitted by the preservation of information in DNA.

The earliest transitions were very occasional – photosynthesis did not appear for around 2 billion years after life began, for example, while neurons arose only around 600 million years ago. Language appeared a mere 100 000 years ago, and has had a strong effect on the biosphere, via human culture and technological development. It seems that complexity can increase the rate at which new channels are discovered, which in turn increases complexity.

The nervous system is a particularly interesting example of a transition in evolution that vastly multiplied the access to channels to larger regions of the phase space of life. Neural processing exponentially multiplies the weight of another role in which *information* plays a role in biology. This is relative information (physical correlation) between the internal state of an organism and the external environment, established via signaling systems and neural representations.

Information –in this sense– is based on physical correlations between the organism and the environment that are exploited by the organism to optimize its behaviours and maximize its survival chances (Dretske 1981, Rovelli 2016b, Kolchinsky and Wolpert 2016 and 2018). (For a discussion on the thermodynamics of this kind of information, see Still, Sivak, Bell and Gavin 2012.). This "relevant" information is the basis of the explosion of the massive elaboration capacity of brains, and, in humans, represents the basis of culture and knowledge themselves. (For attempts to understand brain functioning in ways related to statistical mechanics, see Friston 2010).

Organisms capable of action (for instance as movement) make use of this form of information about the external world (that is, correlations between their internal states and external features of the world) in order to determine which alternative actions to follow. A bacterium can detect a sugar gradient to move towards a region richer in nutrients. In the same way, neuronal processes can encode information of this kind in the formation of synapses and in semi-stable dynamical paths. Nervous systems allow fast storage of meaningful information and the possibility of elaborating it. This is the foundation of memory, and therefore the early origin of experiential time, which depends on memory (Eichenbaum 2017, Rovelli 2018)

Life has found effective ways to maximize its capacity for exploring the space of possibilities. One of these is sexual reproduction. A second is movement, which allows an organism to explore a vastly greater region of the phase space than before - an effect that is multiplied when other organisms also move, opening up possibilities for new selection forces such as competition and predation. Another is mortality, which gets rapidly rid of individual organisms, to leave place for the next generation to put its random DNA novelties to the test. Two very general consequences of all this are the virtual impossibility of knowing the future and the fact that we die. We may not like either of them, or not, but this is how life works.

Like any exploration, life's trial and error exploration of complexity has seen dramatic drawbacks and true catastrophes. Mass extinctions (one of which we are experiencing right now) for instance, or life's novel photosynthetic self-poisoning with oxygen 300 million years ago, which almost wiped out the early biosphere, are examples. Certainly life has found many dead-ends, while progressing towards greater complexity.

The overall stability of the biosphere is obviously remarkable, since it has persisted for a good fourth of the age of the known universe, but this stability has been realized in trials and errors with quite numerous intermediate catastrophes with very many species wiped out in each of them. Random exploration leads to random results.

**XIII.  A RECENT STEP: HUMANITY**

In closure, we turn to the human species; the product of a transitional step in evolution that has further increased the complexity of life's activities. Humans have evolved a cognitive representational capability, namely thinking and language, that allows us to create new correlations across time and space – that is, to create new forms of macroscopic order to funnel entropy into metabolism. This is manifest in many ways. For example, thanks to these tools, the experiential time of our species is enormously dilated, giving us a wide sense of time flow (see Rovelli 2018). We are aware of distant past and can plan far more ahead than any other species. But our peculiar sense of time is thus ultimately rooted in the temporal resilience of the DNA molecules.

Language allows humans to cooperate in learning and planning; the experience of one individual can be propagated to many others to a degree previously unparalleled. Writing, and more recently electronic media, has amplified cultural transmission of information, allowing us to develop technology that has extended our lifespans and our reach across the planet, and beyond. Money, which is a form of information, allowed us to extend simple forms of resource exchange into new realms – ones that extend across time and space and also from concrete goods like food to ephemera such as 'futures'.

The increasingly complex information elaborated by our individual brains, multiplied collectively by the interactions in our cultural space, has had a strong impact on the entire biosphere. The face of the planet can now be changed by something even more evanescent than a random molecular DNA mutation: it can be heavily affected by the result of the flickering neuron's firing in the endlessly variable networks of one or more brains, where synapses are constantly strengthened and weakened by the billions, allowing flows of information and energy that constantly enable metabolism and entropy growth.

If the space of the possible chemical combinations is stratospheric, far larger is the space of the possible synaptic combinations: we have $10^{11}$ neurons with roughly $10^5$ synapses each. If at a given time each neuron can be firing or not, this gives more than astronomical number of $2^{10^{11}}$ as the number of possible brain states. These states correlate themselves with the external environment (via senses), with the past states of affairs (via memory stored in the synaptic connectivity and in the dynamical network processes) and with many other brains (via language and culture), forming a powerful tool to dealing with information and elaborating it. Let us imagine for example that on average a person has a new idea at least once an hour. Then for our species novelty emerges in the mental world at least $10^5$ times more frequently than in the biological one (Pollack A. and R., 2015).

The very fact we can do science and discuss all this, as in this article, is the product of this jump up in complexity allowing fast elaboration of vast amount of information.

In a sense, the huge effects that the neural dynamics underlying our mental states can have on the entire biosphere –thanks to the heavy coupling between different scales that is all over biology– allows us to sidestep more normal biological evolution and strongly shortens the pace of change. In a few millennia, for instance, we have altered the Earth's mammalian biomass to the point that today 90% of the mammalian biomass is either ourselves or species we raise ourselves to kill and eat (Bar-On et al., 2018). Since the Industrial Revolution our species has increased in numbers to $7 \times 10^9$, far above the $\sim 10^5$ individuals in other mammalian species whose individuals are of our size.

From a biological perspective all this sounds like an unusually strong fluctuation that the biosphere may very well have difficulty in sustaining. Complexity is not necessarily unstable, but not necessarily stable either, as we have seen with the several mass extinctions of life that have occurred in earth's past history. The human-specific emergent properties of our species are relatively new, and could easily be unstable.

One may notice that the search for other "intelligent life" in the universe by searching for structure in the electromagnetic signals arriving to us from the cosmos has so far failed. This might simply be because of the vastness of the galaxy, or because of our narrow imagination in figuring out how other forms of "intelligent life" may work. But it could also be because "intelligent life", namely the power of fast elaboration of information, is biologically unstable because it gives itself large powers of self destruction, so that it typically self-destroys in a few centuries, giving too narrow a gap for us to observe similar experiments of nature elsewhere in the galaxy (Pollack, R., 2015)

The chances of bringing about the end of intelligent life by self-destruction seem to be very high, be it via climate change, mass extinction, nuclear war or something else. We do not know whether the very recent life's experiment that our species represents is a dead-end or a step with some future. Reason and far-sighted behaviours might perhaps be able to control these threats, but the call is open, and subject to the way we elaborate information and decide, individually and collectively. For this to happen, humanity must learn to act together politically, putting its common vital interest above the interests of classes, groups, nations and individuals. It is our sense that time is running out for this to learning experience to emerge, and we welcome suggestions and ideas for how to bring it about. Such suggestions would be, if successful, channels to new and larger regions of the phase space of life.

## XIV. SUMMARY

How do we make plausible the persistence, through semi-conservative replication over billions of years, of seemingly improbable informational structures such as the DNA molecules, which govern life and carry the information for large structure formation sustained by metabolism? Even if an extremely fortuitous combination of factors jump-started the process on our minuscule planet, how did the process survive four billion years of catastrophes? In this paper we have offered a statistical interpretation meant to dispel the apparent cogency of these puzzles.

Summarizing: assume that the initial macro-state has low entropy. Consider the fact that low entropy photons downpour on Earth, providing an endless source of negative entropy. Because of the nonlinear form of the physical laws, and because of the extraordinary multiplying power of combinatorics, the energetic landscape of physical processes is extremely complex. Microstates that belong to a low entropy macro-state get typically trapped into metastable states. Entropy does not grow to its maximum. But channels can be open, through which a microstate can move up to higher entropy. If it can, it does it, simply for probabilistic reasons, those underpinning the second law. Channels are opened by the existence of structured, ordered, configurations, because these allow processes that would not happen in their absence. Particularly efficient in raising entropy are structures that act for long times, therefore paths that have such structures in time, namely to long-range correlations in time, namely to preservation of information for long times, are entropically favored.

If this statistical interpretation holds, the following is the way of reinterpreting characteristic life phenomena statistically.

*Metabolism* is the main entropic process happening: it is simply an irreversible phenomenon, directly driven by the second law of thermodynamics. It happens for the same reason a candle burns: because it is entropically favored. From a statistical perspective, it is not metabolism that serves the development and preservation of life, but, in a sense, is the other way around: it is the structured and ordered form of life which allows metabolism to happen. The fire of the candle is permitted by the structure of a candle: it is not the candle that is permitted by its fire.

*Structure* is order. It can be quantified in terms of correlations. Formation of organic molecules is formation of order. The reason for this order is that it opens channels for entropy to grow in the very intricate energetic structure of the state space. There are several types of order in biology: synchronic structural (instantaneous) order internal and external to the living system, and diachronic order, namely order in time. Order is measured by Shannon's relative information.

*Information* can be defined rigorously and quantitatively in terms of physical correlation. It plays a number of different roles:

(a) A key role in biology is played by the information contained in DNA. DNA strands have information about (are correlated with) one another, and about the proteins they encode. Furthermore, thanks to the reproduction allowed by their double-strand nature, their information is carried from one time to another (molecules are correlated across time). The DNA molecule has carried information across time-spans that reach billions of years. The existence of this structure is what allows the long-term efficiency of the entropy growth, and therefore what makes the process entropically favored. Life relies on relative information across very long time spans.

(b) A second kind of information ("relevant information") is that component relevant for survival of the correlations between the internal state of an organism with the external environment (Dretske 1981, Rovelli 2016b, Kolchinsky and Wolpert 2016, 2018). This is the basis of the information massively elaboration by brains, and indirectly the basis of the information forming culture and knowledge themselves.

*Evolution* is slow percolation in an extremely complex energetic landscape. It is the progressive opening of new channels for entropy to grow. It progresses from less-complex to more-complex as increased complexity offers an increased number of new channels for entropy growth.

*Jumps in evolution* are the nodal points wherein the process suddenly discovers new channels for entropy to grow, by finding new forms of (relatively) permanent order allowing this. The existence of these jumps is a direct consequence of the complex structure formed by the metastable configurations in the space of the histories of the dynamical trajectories in complex energetic landscape of carbon chemistry.

*Improbability of life* is a mistaken idea. It comes from the erroneous identification of high entropy with disorder, from the mistake of confusing the probability of an event and the probability of a class of events, from our inability to predict the possible forms for structure to open channels for entropy to grow, and from a naive intuition of thermodynamics that ignores metastable states, channels between these, and the extreme complexity of the energetic landscape of the real physical world.

*Catastrophe* occurs when a system discovers channels into higher-entropy lower-order states. Sometimes the system can escape again, as with recovery from near-extinction; at other times it may not. The future for life on Earth, including human life, is currently uncertain: whether we escape into a higher-order or lower-order part of the phase space depends on what channels we choose.

***

The authors wish to thank Eilam Lehrman, Matteo Pollettini and Matteo Smerlak for their willingness to read and comment on drafts of this paper. They acknowledge support from the Center for Science and Society of the Columbia University (RP), the Wellcome Trust (KJ), and the FQXi Institute (CR). KJ and CR are grateful to the Kavli Foundation for supporting the April 2018 « Kavli Salon: Space, Time and the Brain » meeting, where some of the ideas presented in this paper were initially discussed.